%% file: conference_101719.tex
\documentclass[a4paper,10pt,conference]{IEEEtran}
\IEEEoverridecommandlockouts
\usepackage{cite}
\usepackage{amsmath,amssymb,amsfonts}
\usepackage{algorithmic}
\usepackage{graphicx}
\usepackage{textcomp}
\usepackage{xcolor}
\usepackage{booktabs} 
\usepackage{subfig}
\usepackage{url}
\usepackage{soul}
\usepackage{tikz}
\usetikzlibrary{arrows.meta, positioning}
\def\BibTeX{{\rm B\kern-.05em{\sc i\kern-.025em b}\kern-.08em
    T\kern-.1667em\lower.7ex\hbox{E}\kern-.125emX}}
    
\usepackage[left=1.65cm,right=1.65cm,top=1.93cm,bottom=2.52cm]{geometry}
\begin{document}

\title{Dualband OFDM Delay Estimation for Multi-Target Localization\\
}

\author{\IEEEauthorblockN{Jialun Kou\textsuperscript{*}, Achiel Colpaert\textsuperscript{†,*}, Zhuangzhuang Cui\textsuperscript{*},  Sofie Pollin\textsuperscript{†,*}}
\IEEEauthorblockA{\textit{\textsuperscript{*}Department of Electrical Engineering (ESAT), KU Leuven, Belgium } \\
\textit{\textsuperscript{†}imec, Kapeldreef 75, 3001 Leuven, Belgium}\\
Email: \{jialun.kou, zhuangzhuang.cui, sofie.pollin\}@kuleuven.be,
achiel.colpaert@imec.be}
}


\maketitle

\input{Context/Abstract}

\input{Context/Introduction}
\input{Context/Signal_model}

\input{Context/Multi-band_Window}
\input{Context/Clean}
\input{Context/Simulation}

\input{Context/Conclusion}
\section*{Acknowledgment}
This work is supported by the SUNRISE-6G and MultiX projects, funded by the Smart Networks and Services Joint Undertaking (SNS JU) under the European Union’s Horizon Europe research and innovation programme, with Grant Agreement Nos. 101139257 and 101192521 and by the Flemish Government under Onderzoeksprogramma Artificiele Intelligentie ¨ Vlaanderen (Flanders AI Research). The work of Z. Cui was supported by the Research Foundation – Flanders (FWO), Senior Postdoctoral Fellowship under Grant No. 12AFN26N.

\input{Context/Reference}
\end{document}

%% file: Context/Abstract.tex




\begin{abstract}


Integrated localization and communication systems aim to reuse communication waveforms for simultaneous data transmission and localization, but delay resolution is fundamentally limited by the available bandwidth. In practice, large contiguous bandwidths are difficult to obtain due to hardware constraints and spectrum fragmentation. Aggregating non-contiguous narrow bands can increase the effective frequency span, but a non-contiguous frequency layout introduces challenges such as elevated sidelobes and ambiguity in delay estimation. This paper introduces a point-spread-function (PSF)-centric framework for dual-band OFDM delay estimation. We model the observed delay profile as the convolution of the true target response with a PSF determined by the dual-band subcarrier selection pattern, explicitly linking band configuration to resolution and ambiguity. To suppress PSF-induced artifacts, we adapt the RELAX algorithm for dual-band multi-target delay estimation. Simulations demonstrate improved robustness and accuracy in dual-band scenarios, supporting ILC under fragmented spectrum.

\end{abstract}

\begin{IEEEkeywords}
Delay estimation, dual-band OFDM, point spread function, RELAX.
\end{IEEEkeywords}

%% file: Context/Introduction.tex
\section{Introduction}

The sixth-generation (6G) wireless networks are expected to support both high-throughput communication and high-precision localization. To meet these dual demands, integrated localization and communication (ILC) has emerged as a key paradigm that reuses communication waveforms for localization tasks~\cite{JCAS}. In many ILC applications, delay estimation (or time-of-arrival (TOA) estimation) is fundamental, and its achievable resolution is primarily governed by the available signal bandwidth. Millimeter-wave radars can leverage large contiguous bandwidths to enable fine delay resolution~\cite{PMCW}. However, practical communication systems often have limited access to wide contiguous bandwidth due to hardware constraints and regulatory spectrum allocation. In particular, systems operating below 7~GHz, such as cellular networks and Wi-Fi, typically provide only tens of MHz per band, which limits delay resolution.

To improve delay estimation performance under bandwidth scarcity, a promising approach is to aggregate multiple non-contiguous narrow subbands. Recent advancements in IEEE 802.11be (Wi-Fi~7) introduce multi-link operation (MLO), enabling devices to establish simultaneous connections across the 2.4~GHz, 5~GHz, and 6~GHz bands~\cite{WIFI7}. This capability allows a receiver to combine measurements from separated subbands and synthesize a larger effective frequency span. Nevertheless, a fragmented (non-contiguous) frequency layout is not equivalent to a truly contiguous band, and it introduces additional challenges for delay estimation, including sidelobe-related ambiguity and the presence of spurious peaks in the delay profile.

Several approaches have been proposed to address multiband delay estimation from non-contiguous measurements~\cite{hisac,vasisht2016decimeter,kazaz2022delay,kazaz2019multiresolution,noschese2021multiband}. One line of work exploits sparsity in the delay domain and reconstructs multipath delays and amplitudes using sparse recovery techniques~\cite{hisac,vasisht2016decimeter}. For example, HiSAC~\cite{hisac} performs sparse reconstruction from fragmented spectrum measurements using orthogonal matching pursuit (OMP), enabling joint processing across separated subbands. Related formulations based on $\ell_1$-regularized optimization have also been investigated for multiband ranging and localization~\cite{vasisht2016decimeter}. Another line of work focuses on parametric super-resolution based on subspace structures across subbands~\cite{kazaz2022delay,kazaz2019multiresolution}. By leveraging shift-invariance relations in the frequency-domain measurements, these methods provide high-resolution delay estimation from channel state information (CSI) collected over multiple bands. In addition, likelihood-based iterative refinement methods have been explored for fusing multiband observations. For instance, the space-alternating generalized expectation-maximization (SAGE) framework has been extended to multiband TOA estimation, where channel parameters are iteratively refined by combining measurements across bands~\cite{noschese2021multiband}. Collectively, these studies demonstrate the potential of multiband operation to improve delay estimation performance under fragmented spectrum.

Despite the progress above, it remains valuable to explicitly link multiband frequency layouts to delay resolution and ambiguity in a unified manner. In this work, we describe the multiband sampling pattern using a binary subcarrier selection function (SCF) and adopt a point-spread-function (PSF)-centric formulation, where the estimated OFDM delay profile is modeled as the convolution of the true target response with an SCF-induced PSF. We focus on dual-band configurations as a representative multiband setting that highlights the key effects of fragmented spectrum on delay estimation.
This formulation naturally motivates deconvolution-based processing~\cite{clean1,clean2,relax} , and we employ RELAX as an efficient iterative deconvolution method for multi-target delay estimation~\cite{relax}.

The main contributions of this paper are summarized as follows:
\begin{itemize}
\item \textbf{PSF characterization.} We derive an analytical characterization of the dual-band PSF and quantify how the inter-band gap determines the mainlobe width and sidelobe structure.
\item \textbf{RELAX-based dual-band delay estimation.} We adapt RELAX to dual-band OFDM measurements and develop a multi-target delay estimator.
\item \textbf{Design guidelines.} We provide practical guidelines for selecting the inter-band gap to balance delay resolution, sidelobe-related ambiguity, and noise robustness.
\end{itemize}

The outline of this paper is as follows: Section~\ref{sec:signal_model} introduces the multiband signal model and PSF characterization. Section~\ref{sec:Impact_dualband} analyzes the impact of the dual-band PSF on delay estimation. Section~\ref{section:intro_clean} presents the RELAX algorithm and its application to dual-band delay estimation. Finally, Section~\ref{sec:simulation_results} presents simulation results and performance evaluations.

%% file: Context/Signal_model.tex
\section{Signal Model}
\label{sec:signal_model}

Consider an OFDM system defined on a $K$-point frequency grid spanning a total frequency span $B$, with subcarrier spacing $\Delta f = B/K$. 
Let $\mathcal{S}=\{k_0,\dots,k_{M-1}\}\subseteq\{0,\dots,K-1\}$ denote the set of active subcarrier indices, with $M=|\mathcal{S}|$. 
We define the binary subcarrier selection function (SCF)
\begin{align}
W[k] \triangleq 
\begin{cases}
1, & k\in\mathcal{S}\\
0, & \text{otherwise}
\end{cases}, \qquad k=0,1,\dots,K-1.
\label{eq:scf_def}
\end{align}

The noiseless channel frequency response (CFR) on the $K$-point grid is modeled as
\begin{align}
H[k] = W[k]\sum_{\ell=0}^{L-1}\alpha_{\ell} 
e^{-\mathrm{j}2\pi k\Delta f\,\tau_{\ell}},
\label{eq:CFR}
\end{align}
where $L$ is the number of targets, and $\alpha_\ell$ and $\tau_\ell$ denote the complex amplitude and propagation delay of the $\ell$-th target, respectively.

To estimate delays, we form the delay profile via an inverse discrete Fourier transform (IDFT) over the selected subcarriers:
\begin{align}
h(\tau) =\frac{1}{K}\sum_{k=0}^{K-1} H[k]\,
e^{\mathrm{j}2\pi k\Delta f\,\tau}.
\label{eq:delay_profile_cont}
\end{align}
Substituting \eqref{eq:CFR} into \eqref{eq:delay_profile_cont} yields
\begin{align}
h(\tau)
=
\sum_{\ell=0}^{L-1}\alpha_{\ell}\,p(\tau-\tau_{\ell}),
\label{eq:delay_profile_psf_cont}
\end{align}
where the PSF induced by the SCF $W[k]$ is defined as
\begin{align}
p(\tau)
=
\frac{1}{K}\sum_{k=0}^{K-1} W[k]\,
e^{\mathrm{j}2\pi k\Delta f\,\tau}.
\label{eq:psf_def_cont}
\end{align}

Equation~\eqref{eq:delay_profile_psf_cont} shows that the multiband delay profile can be written as the convolution of a multi-target delay response with a PSF $p(\tau)$ determined by the SCF $W[k]$.  For a full-band system ($W[k]\equiv 1$), $p(\tau)$ reduces to the Dirichlet kernel associated with a contiguous band. Under fragmented spectrum, however, $p(\tau)$ depends on the selected subcarriers and typically exhibits elevated sidelobes, which increases ambiguity in delay estimation. This convolutional formulation enables deconvolution-oriented processing to suppress PSF-induced artifacts. In Section~\ref{section:intro_clean}, we employ RELAX as an efficient iterative method to estimate multi-target delays and complex amplitudes from dual-band measurements~\cite{relax}.

%% file: Context/Multi-band_Window.tex
\section{Impact of the dual-band point spread function} \label{sec:Impact_dualband}

\begin{figure*}[!t]
\centering
\subfloat[SCF, $f_{\mathrm{gap}}=40$~MHz\label{fig:scf_gap40}]
{\includegraphics[width=0.3\linewidth]{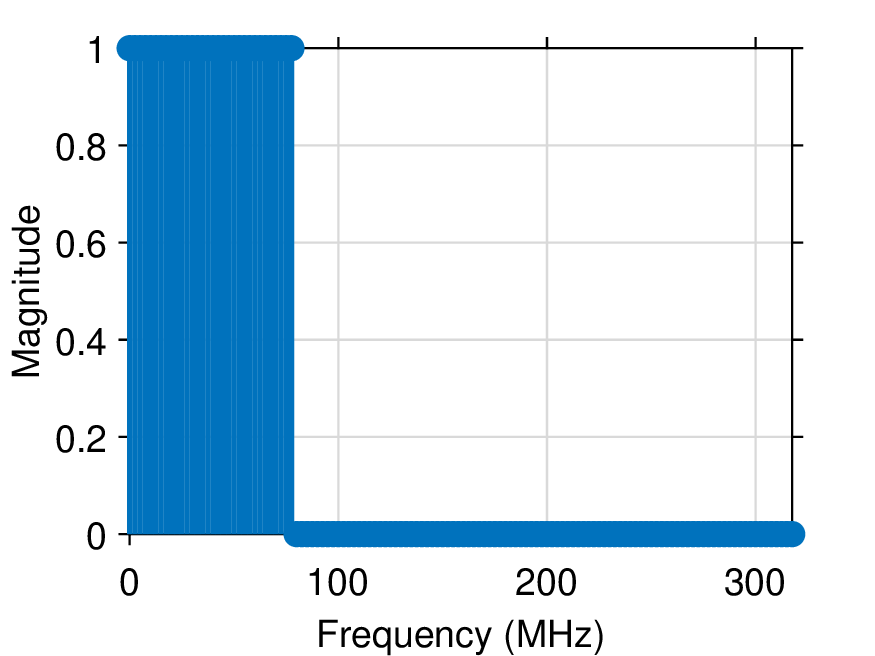}}
\hfill
\subfloat[SCF, $f_{\mathrm{gap}}=120$~MHz\label{fig:scf_gap120}]
{\includegraphics[width=0.3\linewidth]{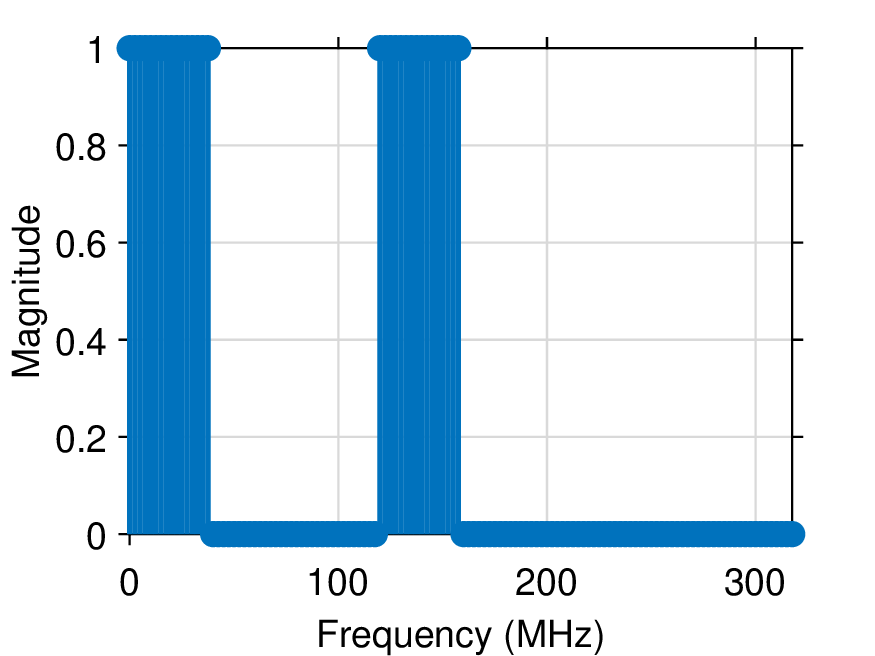}}
\hfill
\subfloat[SCF, $f_{\mathrm{gap}}=280$~MHz\label{fig:scf_gap280}]
{\includegraphics[width=0.3\linewidth]{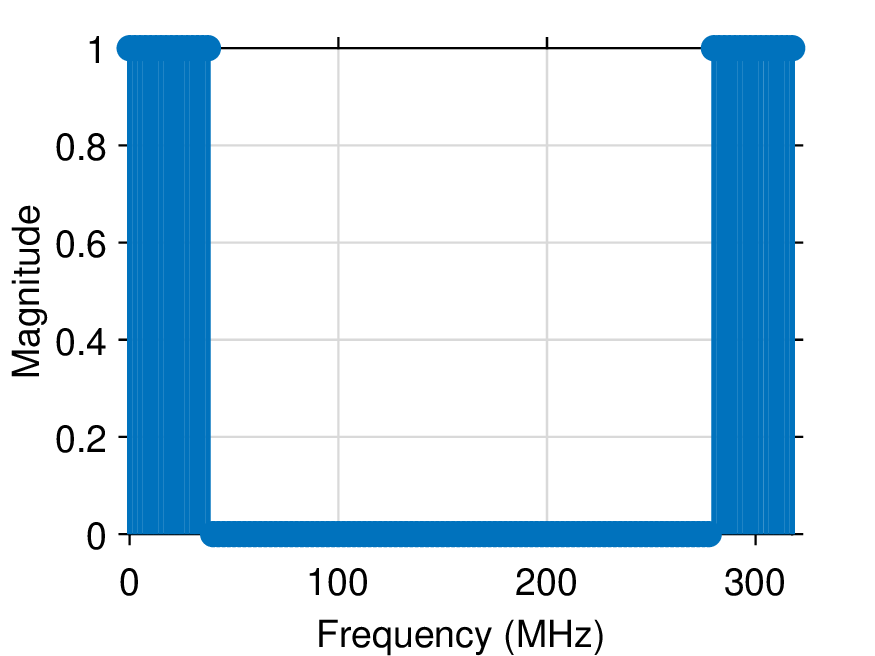}}
\\[1ex]
\subfloat[PSF, $f_{\mathrm{gap}}=40$~MHz\label{fig:psf_gap40}]
{\includegraphics[width=0.3\linewidth]{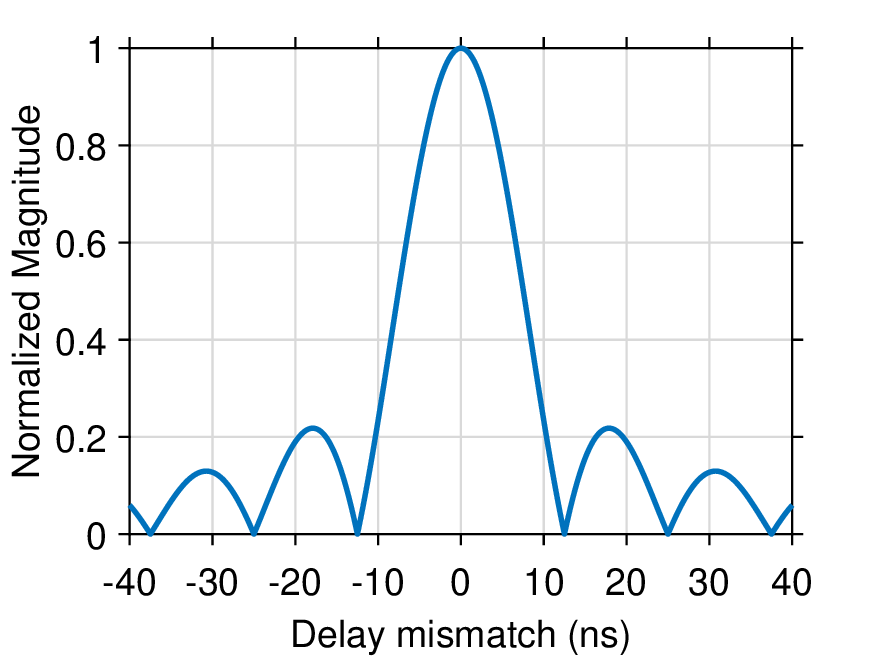}}
\hfill
\subfloat[PSF, $f_{\mathrm{gap}}=120$~MHz\label{fig:psf_gap120}]
{\includegraphics[width=0.3\linewidth]{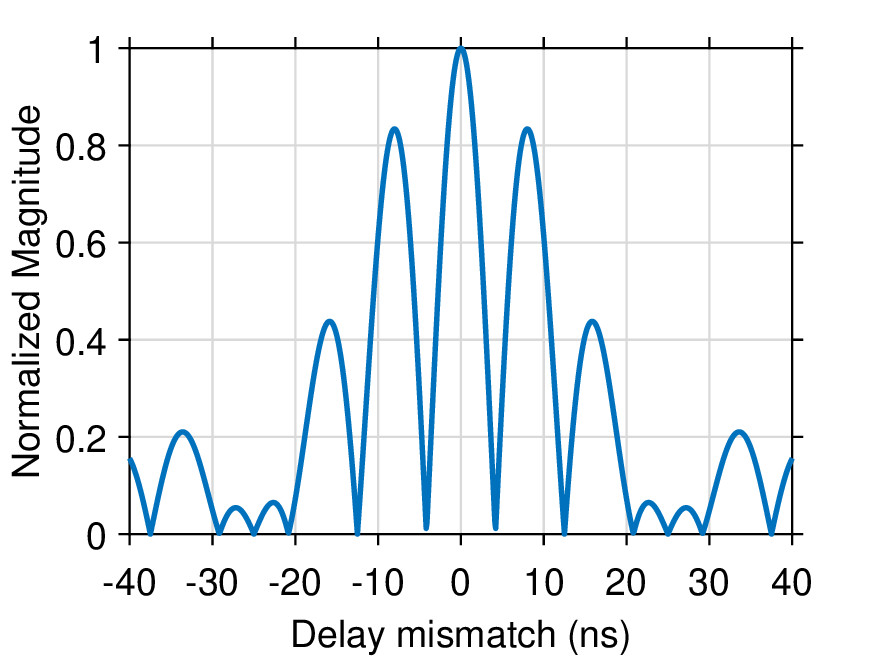}}
\hfill
\subfloat[PSF, $f_{\mathrm{gap}}=280$~MHz\label{fig:psf_gap280}]
{\includegraphics[width=0.3\linewidth]{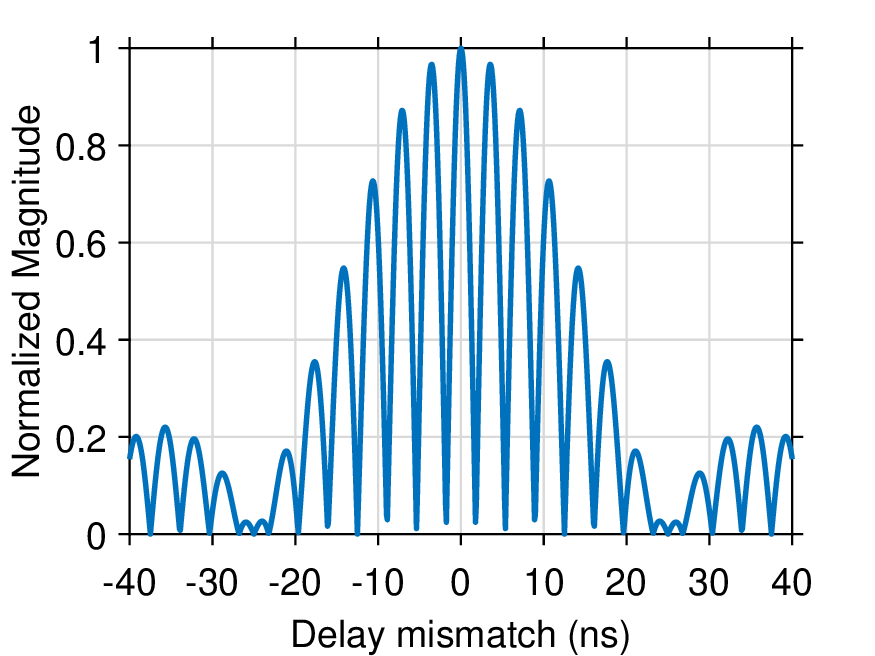}}
\caption{Subcarrier selection functions (SCFs) (top row) and the corresponding point-spread functions (PSFs) (bottom row) for different center-frequency gaps $f_{\mathrm{gap}}$. Each subband has bandwidth $B_{\mathrm{sub}}=40$~MHz.}
\label{fig:PSF}
\end{figure*}

Building on the  PSF definition in \eqref{eq:psf_def_cont}, we now specialize to a dual-band setting with two active subbands, each of bandwidth $B_{\mathrm{sub}}=40$~MHz. We define $f_{\mathrm{gap}}$ as the center-to-center frequency separation between the two subbands, as illustrated in Fig.~\ref{fig:PSF}.

\subsubsection{Adjacent subbands ($f_{\mathrm{gap}}=B_{\mathrm{sub}}$)}
When the two 40~MHz subbands are adjacent (i.e., there is no empty frequency gap), the center-to-center separation equals one subband bandwidth, as shown in Fig.~\ref{fig:scf_gap40}. In this case, the selected subcarriers form a single contiguous block spanning $2B_{\mathrm{sub}}=80$~MHz. Therefore, $p(\tau)$ reduces to the classical Dirichlet-kernel response of a contiguous band, as observed in Fig.~\ref{fig:psf_gap40}. The nominal delay resolution is therefore determined by the total occupied bandwidth $2B_{\mathrm{sub}}$.

\subsubsection{Separated subbands ($f_{\mathrm{gap}}>B_{\mathrm{sub}}$)}
When the two subbands are separated, the active indices consist of two disjoint contiguous blocks. Let each subband contain $N=B_{\mathrm{sub}}/\Delta f$ active subcarriers, and define the center-to-center separation in subcarrier indices as \(g \triangleq f_{\mathrm{gap}}/ \Delta f,\) where $g\geq N$ for $f_{\mathrm{gap}}\geq B_{\mathrm{sub}}$. Without loss of generality, the active set can be written as
\begin{align}
\mathcal{S}
=
\left\{0,1,\ldots,N-1\right\}
\cup
\left\{g, g+1,\ldots,g+N-1\right\}.
\label{eq:active_set_dualband}
\end{align}
Substituting this selection pattern into \eqref{eq:psf_def_cont} yields
\begin{align}
p(\tau)
&=
\frac{1}{K}\sum_{m=0}^{N-1} e^{\mathrm{j}2\pi m\Delta f\,\tau}
+
\frac{1}{K}\sum_{m=0}^{N-1} e^{\mathrm{j}2\pi (m+g)\Delta f\,\tau}
\nonumber\\
&=
\Bigl(1+e^{\mathrm{j}2\pi f_{\mathrm{gap}}\tau}\Bigr)\cdot
\frac{1}{K}\sum_{m=0}^{N-1} e^{\mathrm{j}2\pi m\Delta f\,\tau}.
\label{eq:psf_dualband_cont}
\end{align}
Equation~\eqref{eq:psf_dualband_cont} shows that the dual-band PSF consists of a single-subband Dirichlet-like envelope multiplied by a cosine-like modulation term. Specifically,
\begin{align}
1+e^{\mathrm{j}2\pi f_{\mathrm{gap}}\tau}
=
2e^{\mathrm{j}\pi f_{\mathrm{gap}}\tau}\cos(\pi f_{\mathrm{gap}}\tau),
\end{align}
so increasing $f_{\mathrm{gap}}$ results in faster oscillations of the cosine-like term in the delay domain. As shown in Fig.~\ref{fig:psf_gap120} and Fig.~\ref{fig:psf_gap280}, larger $f_{\mathrm{gap}}$ yields narrower peaks, which can improve discrimination between closely spaced targets at high SNR. Meanwhile, it also produces more closely spaced sidelobe peaks around the dominant maximum, increasing ambiguity and the risk of selecting a sidelobe peak instead of the true peak at low SNR.

\begin{figure}[t]
\centering
\includegraphics[width=0.9\linewidth]{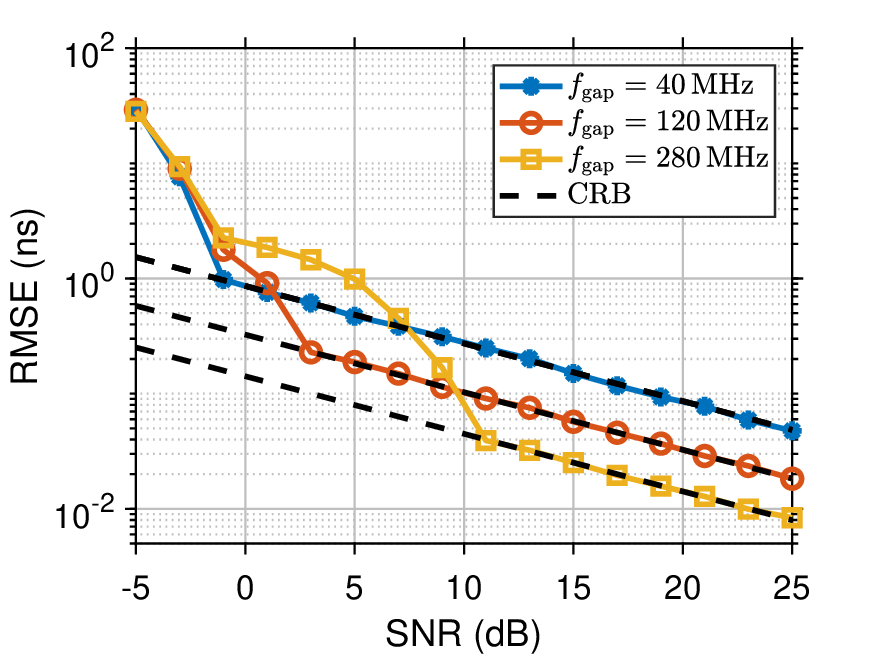}
\caption{Single-target delay RMSE versus SNR for different center-frequency gaps $f_{\mathrm{gap}}$, together with the corresponding Cram\'er--Rao bounds (CRBs).}
\label{fig:impact_of_frequencygap}
\end{figure}

Fig.~\ref{fig:impact_of_frequencygap} reports the single-target delay RMSE versus SNR for several values of $f_{\mathrm{gap}}$. As $f_{\mathrm{gap}}$ increases, the maximum-likelihood estimator (MLE) achieves a lower high-SNR error floor, indicating improved asymptotic accuracy. However, the SNR threshold at which the MLE approaches the CRB increases with $f_{\mathrm{gap}}$. This trend is consistent with the PSF structure: larger $f_{\mathrm{gap}}$ leads to stronger and more closely spaced sidelobes, which increases the probability of incorrect peak selection at low SNR. In summary, increasing the center-frequency gap can improve high-SNR accuracy but also increases sensitivity to noise. Note that this discussion is based on a single-target scenario; with multiple closely spaced targets, sidelobe overlap can further degrade estimation performance.

%% file: Context/Clean.tex
\section{RELAX Algorithm for PSF Deconvolution}
\label{section:intro_clean}

To mitigate sidelobe-induced artifacts in dual-band delay estimation, we employ the RELAX algorithm to estimate target delays and complex amplitudes directly from fragmented-spectrum measurements. We consider the dual-band active subcarrier set $\mathcal{S}$ defined in \eqref{eq:active_set_dualband}. 
We denote by $\mathbf{y}\in\mathbb{C}^{M}$ the measurement vector obtained by stacking the observed CFR samples on the active subcarriers indexed by $\mathcal{S}$. The corresponding delay steering vector is defined as
\begin{align}
\mathbf{a}_{\mathcal{S}}(\tau)
=
[\,e^{-\mathrm{j}2\pi k\Delta f\,\tau}\,]_{k\in\mathcal{S}}
\in\mathbb{C}^{M},
\label{eq:steer_dualband}
\end{align}
where $\mathrm{j}$ denotes the imaginary unit. The measurement model is
\begin{align}
\mathbf{y} = \sum_{\ell=1}^{L}\alpha_\ell\,\mathbf{a}_{\mathcal{S}}(\tau_\ell) + \mathbf{n},
\label{eq:relax_meas_model}
\end{align}
where $\mathbf{n}\sim\mathcal{CN}(\mathbf{0},\sigma^2\mathbf{I}_{M})$ denotes additive white Gaussian noise (AWGN) on the observed subcarriers.

Under the AWGN assumption, estimating $\{(\alpha_\ell,\tau_\ell)\}_{\ell=1}^{L}$ corresponds to the maximum-likelihood (ML) problem
\begin{align}
(\widehat{\boldsymbol{\alpha}},\widehat{\boldsymbol{\tau}})
&=\arg\min_{\{\alpha_\ell,\tau_\ell\}_{\ell=1}^{L}}
\left\|
\mathbf{y}-\sum_{\ell=1}^{L}\alpha_\ell\,\mathbf{a}_{\mathcal{S}}(\tau_\ell)
\right\|_2^{2}.
\label{eq:relax_obj_align}
\end{align}
A direct solution requires a costly $2L$-dimensional search. RELAX avoids this by alternating single-target updates within outer refinement cycles.

\subsection{RELAX procedure for delay estimation}
We perform a coarse grid search over a discretized delay set $\mathcal{T}$ and then iteratively refine the target delays and amplitudes using residual-based updates.

\begin{enumerate}
\item \textbf{Initialization:}
Set $J \leftarrow 0$ and initialize the residual as
\begin{align}
\mathbf{r} \leftarrow \mathbf{y}.
\end{align}
Choose a maximum number of targets $L_{\max}$ and a stopping threshold $\varepsilon$.

\item \textbf{Target acquisition (add one target):}
Increase the model order $J \leftarrow J+1$. A new delay is acquired by matched-filter (correlation) search:
\begin{align}
\tau_J
&\in \operatorname*{arg\,max}_{\tau\in\mathcal{T}}
\bigl|\mathbf{a}_{\mathcal{S}}^{H}(\tau)\,\mathbf{r}\bigr|,
\label{eq:acquire_tau}\\
\alpha_J
&=\frac{\mathbf{a}_{\mathcal{S}}^{H}(\tau_J)\,\mathbf{r}}
        {\|\mathbf{a}_{\mathcal{S}}(\tau_J)\|_2^{2}}.
\label{eq:acquire_alpha}
\end{align}
This step selects the delay that best explains the current residual and then updates its amplitude by least squares.

\item \textbf{RELAX refinement cycles :}
Repeat the following updates for $\ell=1,\ldots,J$:
\begin{enumerate}
\item \textbf{Per-target residual:}
\begin{align}
\mathbf{r}_{\ell}
&=\mathbf{y}
-\sum_{\substack{p=1\\ p\neq \ell}}^{J}
\alpha_p\,\mathbf{a}_{\mathcal{S}}(\tau_p).
\label{eq:per_comp_resid}
\end{align}
This residual removes the contribution of all targets except the $\ell$th one.
\item \textbf{Single-target update:}
\begin{align}
\tau_\ell
&\in \operatorname*{arg\,max}_{\tau\in\mathcal{T}}
\bigl|\mathbf{a}_{\mathcal{S}}^{H}(\tau)\,\mathbf{r}_{\ell}\bigr|,
\label{eq:update_tau}\\
\alpha_\ell
&=\frac{\mathbf{a}_{\mathcal{S}}^{H}(\tau_\ell)\,\mathbf{r}_{\ell}}
        {\|\mathbf{a}_{\mathcal{S}}(\tau_\ell)\|_2^{2}}.
\label{eq:update_alpha}
\end{align}
This step refines $\tau_\ell$ via correlation maximization and updates $\alpha_\ell$ via least squares.

\end{enumerate}

\item \textbf{Residual refresh:}
After updating all $J$ targets, recompute the global residual as
\begin{align}
\mathbf{r}
=\mathbf{y}-\sum_{p=1}^{J}\alpha_p\,\mathbf{a}_{\mathcal{S}}(\tau_p).
\label{eq:resid_refresh_correct}
\end{align}

\item \textbf{Stopping criterion:}
If $\|\mathbf{r}\|_2^{2}\le\varepsilon$ or $J=L_{\max}$, stop and output $\{(\alpha_\ell,\tau_\ell)\}_{\ell=1}^{J}$.
Otherwise, return to the target acquisition step to add the next target.
\end{enumerate}

\subsection{Reconstruction}
After convergence, RELAX outputs the estimates $\{(\alpha_\ell,\tau_\ell)\}_{\ell=1}^{J}$. 
To reconstruct the corresponding full-band response, we use the full-band steering vector
\begin{align}
\mathbf{b}(\tau)
\triangleq
\bigl[\,e^{-\mathrm{j}2\pi 0\Delta f\,\tau},\ldots,
       e^{-\mathrm{j}2\pi (K{-}1)\Delta f\,\tau}\,\bigr]^T,
\end{align}
and reconstruct
\begin{align}
\widehat{\mathcal{R}}_{\mathrm{full}}=\sum_{\ell=1}^{J}\alpha_\ell\,\mathbf{b}(\tau_\ell),
\end{align}
which can be transformed to the delay domain via an IDFT.

%% file: Context/Simulation.tex
\section{Simulation results}\label{sec:simulation_results}

\subsection{Simulation setup}
To evaluate the proposed dual-band OFDM delay estimation framework, simulations are conducted using the parameters summarized in Table~\ref{tab:radar_params}. We consider a total 320~MHz frequency grid partitioned into eight non-overlapping 40~MHz subbands, each containing 128 subcarriers. In this work, two subbands are activated for delay estimation, resulting in 256 active subcarriers (80~MHz occupied bandwidth). Different dual-band configurations are obtained by varying the center-frequency gap $f_{\mathrm{gap}}$ between the two active subbands.

The scene comprises $L=3$ static targets with fixed delays
\(
\{\tau_\ell\}_{\ell=1}^{L}=\{66,\,100,\,133\}\,\text{ns}.
\)
For each trial, the complex gains $\{\alpha_\ell\}$ are drawn as circularly symmetric complex Gaussian random variables, so that the magnitudes are Rayleigh distributed. Noise is modeled as AWGN, and the SNR is swept from $-10$~dB to $35$~dB.

\begin{table}[!t]
\renewcommand{\arraystretch}{1.2}
\caption{OFDM parameters used in the simulation}
\label{tab:radar_params}
\centering
\begin{tabular}{ll}
\toprule
\textbf{Parameter} & \textbf{Value} \\
\midrule
Carrier frequency                          & 5.2~GHz \\
Total number of subcarriers                & 1024 \\
Number of selected subcarriers             & 256 \\
Total spanned bandwidth                    & 320~MHz \\
Occupied bandwidth                         & 80~MHz \\
Subcarrier spacing                         & 312.5~kHz \\
\bottomrule
\end{tabular}
\end{table}

\subsection{Performance analysis}\label{senction:psf_rangeestimation}

\begin{figure}[t]
\centering
\includegraphics[width=0.9\linewidth]{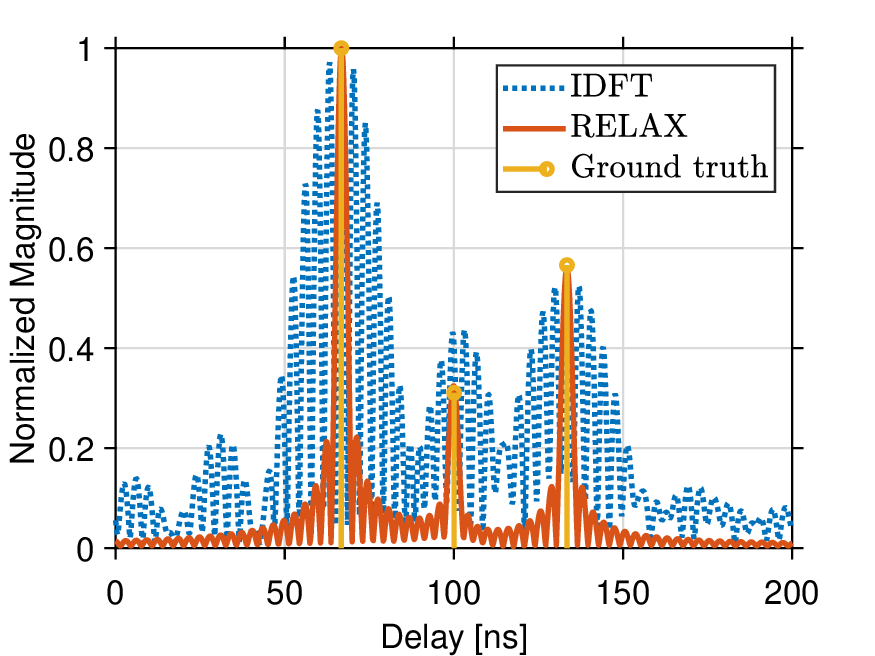}
\caption{Delay profile comparison: IDFT-based dual-band profile and RELAX reconstruction.}
\label{fig:single_target}
\end{figure}

\begin{figure}[t]
\centering
\includegraphics[width=0.9\linewidth]{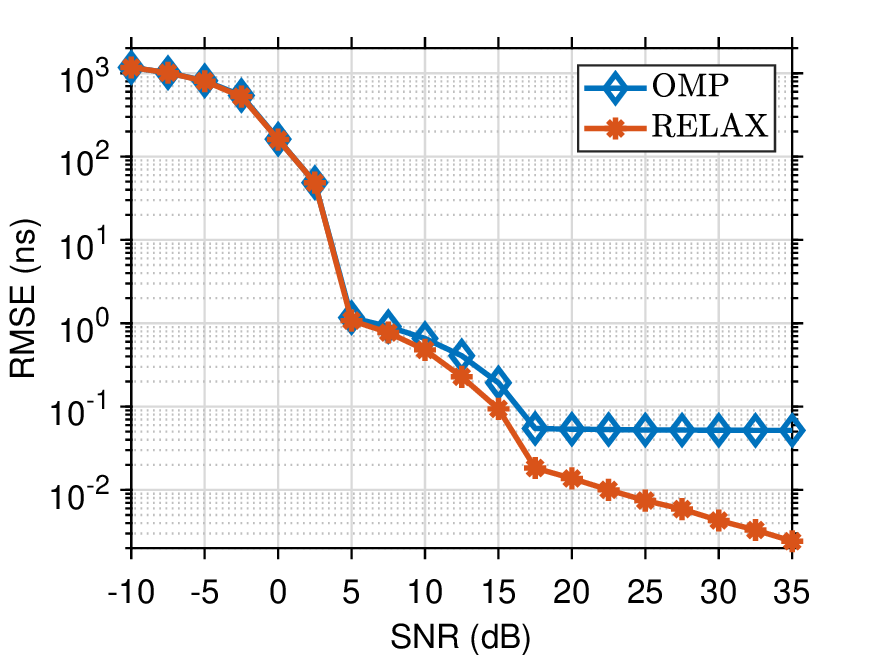}
\caption{Delay RMSE comparison of OMP and RELAX.}
\label{fig:diff_algorithm}
\end{figure}

\begin{figure}[t]
\centering
\includegraphics[width=0.9\linewidth]{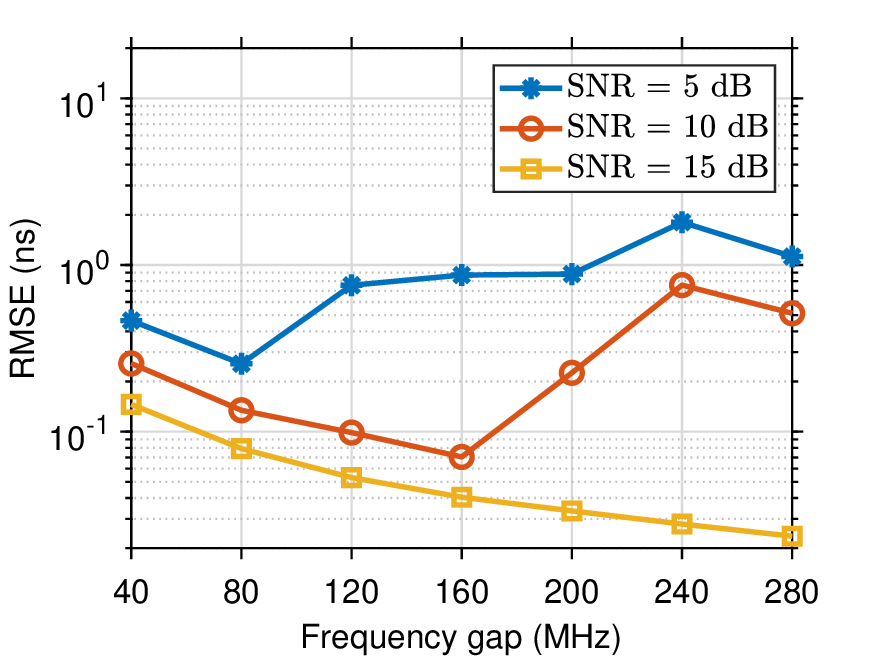}
\caption{Impact of center-frequency gap $f_{\mathrm{gap}}$ on RELAX performance.}
\label{fig:multi_target}
\end{figure}

Fig.~\ref{fig:single_target} compares two delay profiles obtained from the same dual-band CFR using different processing methods: a direct IDFT-based profile and the RELAX-based reconstruction, for the dual-band configuration in Fig.~\ref{fig:scf_gap280}.

The IDFT-based profile is severely corrupted by PSF sidelobes (see Fig.~\ref{fig:psf_gap280}), which can create spurious peaks and obscure the true target locations. In contrast, RELAX suppresses most sidelobe-induced artifacts and yields a delay profile that better reflects the true target delays.
Next, we compare RELAX with the OMP-based method in~\cite{hisac}. We quantify delay estimation accuracy using the delay root-mean-square error (RMSE),
\begin{align}
e_{\tau}^{\mathrm{RMSE}}
\triangleq
\sqrt{
\mathbb{E}\!\left[
\frac{1}{L}\sum_{\ell=1}^{L}
\left(\hat{\tau}_{\ell}-\tau_{\ell}\right)^2
\right]},
\label{eq:rmse_delay}
\end{align}
where the expectation is approximated by averaging over 1000 Monte Carlo trials.

As shown in Fig.~\ref{fig:diff_algorithm}, RELAX achieves a lower RMSE than OMP across the SNR range. OMP is a greedy procedure that detects targets sequentially; consequently, errors in early detections can propagate and degrade subsequent estimates, especially when targets are closely spaced or sidelobes are strong. In contrast, after adding a new component, RELAX cyclically refines the delays and amplitudes of all detected targets via residual-based updates, which reduces mutual interference at the cost of increased computational complexity.

Fig.~\ref{fig:multi_target} reports RELAX performance versus the center-frequency gap $f_{\mathrm{gap}}$ at different SNRs. In this experiment, the first subband is fixed while the second subband is shifted to realize different gaps. At high SNR (e.g., 15~dB), the RMSE decreases as the gap increases, indicating improved delay resolution due to a larger frequency separation. At low SNR (e.g., 5~dB), the RMSE becomes non-monotonic: it initially decreases for moderate gaps but increases again for large gaps. This behavior is consistent with the PSF structure. Larger gaps lead to stronger and more closely spaced sidelobes with reduced contrast relative to the main peak, which increases the risk of selecting a sidelobe peak at low SNR. Overall, larger $f_{\mathrm{gap}}$ improves high-SNR accuracy but makes the estimator more sensitive to noise at low SNR.

%% file: Context/Conclusion.tex
\section{Conclusion}
This paper formulated multi-band OFDM delay estimation using a PSF-centric model, clarified how the inter-band gap affects resolution and sidelobe behavior, and adapted RELAX to exploit this structure to reduce PSF-induced sidelobe artifacts. The analysis provides a guidance for selecting the inter-band gap to balance resolution against sidelobe-induced ambiguity and noise sensitivity.
Future work includes robust weighting for sidelobe suppression, extension to joint delay–Doppler estimation in mobility, and over-the-air validation in the presence of hardware impairments.
